# Quantum Monte Carlo, time-dependent density functional theory, and density functional theory calculations of diamondoid excitation energies and Stokes shifts

F. Marsusi[1*], J. Sabbaghzadeh[2], and N. D. Drummond[3]

[1]Organic Solar Cell group at National Center for Laser Science and Technology, Tehran, Iran

[2]National Center for Laser Science and Technology, Tehran, Iran

[3]Department of Physics, Lancaster University, Lancaster LA1 4YB, United Kingdom

We have computed the absorption and emission energies and hence Stokes shifts of small diamondoids as a function of size using different theoretical approaches, including density functional theory and quantum Monte Carlo (QMC) calculations. The absorption spectra of these molecules were also investigated by time-dependent density functional theory (TD-DFT) and compared with experiment. We have analyzed the structural distortion and formation of a self-trapped exciton in the excited state, and we have studied the effects of these on the Stokes shift as a function of size. Compared to recent experiments, QMC overestimates the excitation energies by about 0.8(1) eV on average. Benefiting from a cancellation of errors, the optical gaps obtained in DFT calculations with the B3LYP functional are in better agreement with experiment. It is also shown that TD-B3LYP calculations can reproduce most of the features found in the experimental spectra. According to our calculations, the structures of diamondoids in the excited state show a distortion which is hardly noticeable compared to that found for methane. As the number of diamond cages is increased, the distortion mechanism abruptly changes character. We have shown that the Stokes shift is size-dependent and decreases with the number of diamond cages. The rate of decrease in the Stokes shift is on average 0.1 eV per cage for small diamondoids.

**PACS numbers: 78.67.Bf, 73.22.-f, 02.70.Ss**

## I. INTRODUCTION

Diamondoids form a series of carbon nanoclusters (C-NCs) that exhibit both diamond-like properties and nanosize effects. They are therefore predicted to find application in a wide range of nanotechnological systems, especially in electronic and optoelectronic devices. For this reason, diamondoids have been the subject of a large number of experimental and theoretical studies in recent years.[1-3] Advances in theoretical condensed-matter physics and first-principles computational methods in the last two decades have made it possible to develop a comprehensive understanding of the electronic and optical properties of a wide range of materials. A great deal of theoretical effort using different many-body techniques has been devoted to calculating the absorption gaps of diamondoids, which are often referred to as optical gaps (OGs).[3-7] However, the measured energy emitted from these materials and the corresponding emission gap (EG) is also of great technological importance. To our knowledge no previous theoretical study has investigated the Stokes shift—the difference of the absorption and emission gaps— exhibited by these novel materials. We have calculated absorption and emission gaps for methane and four small diamondoids, adamantane ($C_{10}H_{16}$), diamantane ($C_{14}H_{20}$), triamantane ($C_{18}H_{22}$), and tetramantane ($C_{22}H_{28}$), in order to predict the Stokes shifts of these C-NCs. We have used several different theoretical approaches, including density functional theory (DFT) and quantum Monte Carlo (QMC), to understand the interplay between their optical absorption and emission gaps, their structural and optical properties, and to interpret the role played by C-C bonds in their Stokes shifts. We have also applied time-dependent density functional theory (TD-DFT)[8-9] to compute the OGs and to make a comparison between the predicted absorption spectra and recent experimental results.

On the basis of the quantum confinement effect (QCE) model, it would be expected that for increasingly large diamondoids the OGs would become progressively smaller. Unexpectedly, in contrast to the QCE model, TD-DFT predicts a larger OG for diamantane than adamantane. Furthermore, by comparison with adamantane and diamantane, the calculated absorption spectrum of triamantane shows a profound change from a discrete molecular spectrum to a quasicontinuous spectrum. Some similar irregularities have also been observed experimentally.[10] Our DFT calculations were performed using (i) plane-wave basis sets within the local density approximation (LDA), and (ii) Gaussian basis sets using the hybrid B3LYP generalized-gradient-approximation functional,[11-12] as the latter has been shown to reproduce precisely the electronic properties of some Si- and C-diamond-like NCs.[7]

For all of the above-mentioned diamondoids we have used QMC methods to calculate the OGs and EGs. In particular, we have used the diffusion Monte Carlo (DMC) method,[13-14] which is one of the most accurate methods available for solving the many-body Schrödinger equation. The computational effort required to achieve a given accuracy scales as $O(N^3)$, where $N$ is the number of electrons in the system. Fermionic antisymmetry is maintained using the fixed-node approximation.[15-17]

The rest of this paper is organized as follows. In Secs. II and III we explain in detail our computational methods and the range of possible errors that may arise in these schemes. In Sec. IV we present and interpret our results, and compare them with the available experimental data. The similarities and differences between theoretical data and corresponding experiments are explained. The geometric response and signature of a self-trapped exciton (STE) state are also explained in this section. Finally a brief summary and our conclusions are given in Sec. V.

## II. Computational methods

We performed LDA plane-wave pseudopotential calculations using the ABINIT code[18] and B3LYP and TD-B3LYP calculations with Gaussian basis sets using the G98 package.[19] Plane-wave cutoffs of at least 32 a.u. and cubic supercells of side-length 32 to 60 a.u. depending on the size of the molecules were used in our LDA calculations. According to previous work,[7] the Gaussian polarized double diffuse 6-31++G** basis set[20-21] is suitable for studying the energy levels of diamondoids and thus was used for all B3LYP and TD-B3LYP calculations in this study. Dirac-Fock (DF) average relativistic effective-core pseudopotentials were used to represent the ionic cores.[22-23] These pseudopotentials are finite at the origin, and the core-core and core-valence electron correlation terms have been neglected, making them particularly suitable for use in QMC calculations.[24] The effect of the choice of pseudopotentials and functional on the DFT and DMC OGs was also tested. In each case, the optimized geometry in the ground state was calculated with the new functional or pseudopotential. The DMC OG of adamantane obtained using DF pseudopotentials is in better agreement with experiment. When DF pseudopotentials were replaced with LDA norm-conserving Troullier-Martins[25] pseudopotentials, the LDA and DMC OGs of adamantane increased by 0.1 and 0.2(1) eV, respectively. When the PBE functional[26] was replaced by the LDA, the DFT OG of adamantane increased by 0.13 eV, while the DMC OG did not change by a statistically significant amount [increasing from 7.35(8) to 7.37(8) eV]. In all our production DFT and DMC calculations, the atomic geometries in both the ground state and the excited state of each molecule were determined using DFT within the LDA framework and a plane-wave basis set, because the Gaussian all-electron calculations failed to converge when relaxing symmetry-constrained excited-state structures in some cases. Our tests show that using the LDA-optimized geometries imposes only a small error on our B3LYP results, reducing both the OG and EG of adamantane by merely 0.08 eV. For our TD-B3LYP calculations, however, the geometries were optimized by means of B3LYP/6-

31++G(d,p) DFT calculations, while the OG is defined as the first singlet, nonzero, allowed optical transition.

The QMC calculations were performed with the CASINO code[27] using Slater-Jastrow[14] trial wave functions of the form $\psi_T = D^{\uparrow} D^{\downarrow} \exp[J]$, where $D^{\uparrow}$ and $D^{\downarrow}$ are Slater determinants of up- and down-spin orbitals taken from LDA calculations. The exp[$J$] is a Jastrow correlation factor, which includes electron-electron, electron-ion, and electron-electron-ion terms expanded as polynomials in the interparticle distances.[28] Plane-wave basis sets are very expensive for large molecules and hence were not used in our DMC calculations; instead we used a B-spline (blip) basis, consisting of piecewise continuous localized cubic-spline functions centered on a regular grid.[29] The Jastrow factors were optimized using a standard variance-minimization scheme.[30-31] Our DMC calculations were performed at time steps of 0.02 and 0.005 a.u., and the results were extrapolated linearly to zero time step. The target population was set to 600 configurations in each calculation. The ionic cores were represented by the same DF pseudopotentials as they were in the DFT calculations.

The DMC optical gap of each diamondoid was calculated as the difference between the DMC energy of the ground state and an excited state. The excited-state wave function for adamantane was constructed using three different strategies and we investigated the sensitivity of the OG to the choice of excitation. The three strategies considered were as follows. (i) Remove a down-spin electron from the highest occupied molecular orbital (HOMO) and add an up-spin electron to the lowest unoccupied molecular orbital (LUMO) in the DFT calculation, then use the resulting DFT orbitals in a single-determinant trial wave function representing the triplet excited state. (ii) Perform a ground-state DFT calculation to generate a set of orbitals, then construct a singlet excitation using a two-determinant wave function in which a single up-spin electron is promoted from the HOMO to the LUMO in the first determinant and a single down-spin electron is promoted from the HOMO to the LUMO in the second determinant. (iii) Perform a ground-state DFT calculation to generate a set of orbitals, then construct the singlet excited-state wave function as an expansion of all the degenerate first-excited-state determinants, and optimize the determinant coefficients within QMC. The DMC OGs obtained for adamantane using the three strategies are 7.35(8), 7.54(8), and 7.48(8) eV, respectively. Therefore the DMC singlet state is insensitive [to within about 0.1(1) eV] to the method used to construct the excited-state trial wave function. Indeed the singlet state is only 0.1(1) eV higher in energy than the triplet state. For computational simplicity we calculated excited-state energies using strategy (i) in all our production DMC calculations. An LDA singlet excited-state calculation for adamantane was performed by removing a down-spin electron from the HOMO and adding a down-spin electron to the LUMO; consistent with DMC, the resulting energy is only 0.12 eV larger than that of the triplet excited state. The effect of using a backflow transformation[32] to improve the nodal surface in the DMC calculations was also investigated for methane and adamantane. The OG of methane was reduced by 0.11(4) eV, but no significant change was seen for adamantane. Our DMC-predicted OGs for adamantane and diamantane are about 0.26(8) eV smaller than the previous DMC work.[3] The DMC calculations of Ref. [3] were more primitive than those reported here in various regards: they used a single, fixed time step of 0.02 a.u., the excited-state wave function was constructed by replacing the HOMO with the LUMO for down-spin electrons in a single-determinant wave function, and DFT pseudopotentials were used to represent the ionic cores.

## III. Stokes shifts

In general, OGs in NCs are larger than EGs. The recombination of the electron and hole from the relaxed atomic configuration leads to a red-shift (longer wavelength) on the emission lines compared to the absorption lines, which is referred as the Stokes shift. According to the Franck-Condon principle, assuming that the electronic transition is very fast compared to the nuclear motion in the molecule, when an electron-hole pair is created by an optical excitation, the final state will be in approximately the same atomic configuration as the initial state. However, before photon emission, the system can relax into a new configuration with a significant decrease in both symmetry and total energy. A schematic diagram of the relevant electronic energy levels as a function of the atomic positions is shown in Fig. 1(a). As the lower-energy triplet state is optically inactive ($\Delta S=0$ rule), the lowest-energy allowed optical transition evolves the system from the ground state [point A in Fig. 1(a)] into the singlet excited state (point $B_S$) with the creation of an electron–hole pair, which is known as an exciton. Possible excitations into different vibrational states associated with the excited state lead to a broadening of the absorption line. Electrons with opposite spins experience larger repulsive Coulomb interactions than parallel-spin electrons, leading to the singlet excited state being higher in energy than the triplet excited state. Consequently, exciton relaxation may be preceded by switching from the singlet state into the triplet state (point $B_T$). If the time required for exciton recombination is longer than the time needed for exciton relaxation, the decrease in the energy continues with a collective relaxation of all atomic positions of the entire molecule into a new geometry (point C). Finally, exciton recombination and photon emission result in a transition of the system from point C to point D (which is at the same geometry as point C). According to the above description, the total OG, EG, and Stokes shift are defined as:

$$\Delta_{OG} = E_B - E_A \qquad (1)$$
$$\Delta_{EG} = E_C - E_D \qquad (2)$$
$$\Delta E_S = \Delta_{OG} - \Delta_{EG} \qquad (3)$$

where $\Delta_{OG}$, $\Delta_{EG}$, and $\Delta E_S$ are the OG, EG, and Stokes shift, respectively. Due to the need to find the relaxed geometry at point C with all symmetry constraints released, computing the Stokes shift is more elaborate than calculating the OG. As explained in Sec. II, we studied the triplet excited state. Then for each molecule we released all of the symmetry constraints and determined the global minimum energy at point C. We investigated the effect of symmetry relaxation in adamantane within the LDA. Optimizing the geometry without the relaxation of the symmetry constraints causes the EG to be 0.67 eV larger than would otherwise be the case, leading to a significant disagreement with the experimental Stokes shift (0.1 eV from the LDA against 0.7 eV from experiment[33]), so symmetry relaxation in the excited state gives an important contribution to the red-shift of the EGs of diamondoids.

### IV. Results and discussion

### A. Optical gaps

Our LDA, B3LYP, TD-B3LYP, and DMC OGs are summarized in Table I. For comparison, previous DMC results[3] along with available experimental data are also given. Two different experimental methods have been used to measure the optical gaps of small diamondoids. Willey et al.[34] measured the filled and empty electronic states of diamondoids in the condensed phase by

X-ray absorption and soft X-ray emission spectroscopy. Landt *et al.*[10] determined the optical absorption of diamondoids in the gas phase by examining the difference of the measured transmission between a filled and an empty absorption cell. The resulting OGs from these two experiments differ by between 0.35 eV and 0.58 eV, which may originate from the different phases of the diamondoids under study. It is more appropriate to compare our theoretical results for isolated molecules with data from gas-phase measurements, so we compare our results with those of Landt *et al*.

All the methods used in this work show nearly the same trend for the OG as a function of size: the OG falls off as the number of cages increases. However, from adamantane to diamantane an irregularity is observed in both the experimental data[10] and the theoretical results, with TD-B3LYP predicting an even larger OG for diamantane than adamantane. The LDA underestimates the experimental gaps, as one would expect, by between 0.73 and 0.97 eV. As shown in Table I, the Hartree-Fock (HF) exchange which is included in the B3LYP functional makes a significant improvement to the DFT OGs of small diamondoids. As can be seen in Table I and Fig. 2, the TD-B3LYP OGs and absorption spectra, the relative intensity of the absorption peaks, and the corresponding wavelengths of the simulated spectra coincide well with the experimental results.

According to Ref. [10], between adamantane and tetramantane a noticeable transition from molecular-like excitations to quasicontinuous spectra takes place. This feature is also found in our theoretical results. The spectrum of adamantane, with a single cage, demonstrates various sharp peaks. Diamantane, with one more cage, exhibits similar features. With the addition of a third cage in triamantane, the sharp peaks move close together, leading to a change in the absorption line from a series of separate peaks into a quasicontinuous feature in the calculated spectrum. The effect of QC is expected to make the OG of diamantane smaller than that of adamantane, but according to Ref. [10], the OG of diamantane is only 0.09 eV smaller than that of adamantane. The shifts in the OG of diamantane relative to adamantane predicted by our LDA, DMC, and B3LYP calculations are 0.33 eV, 0.3(1) eV, and 0.18 eV, respectively. On the other hand, according to our TD-B3LYP calculations, the first two dipole moment transitions from the HOMO to the LUMO, and the HOMO-1 to the LUMO levels of diamantane have zero oscillator strength according to the parity selection rule ($E_g$ to $A1_g$ transition); the first allowed electronic transition with nonzero oscillator strength is from the HOMO to the LUMO+1 at 6.51 eV, which is 0.1 eV larger than the OG of adamantane. The effect of the symmetry of the orbitals on the OG was not included in either this or previous DFT and DMC theoretical efforts. The same irregularity is reported in TD-PBE0 calculations,[35] which indicate that the OG of diamantane is larger than that of adamantane. TD-B3LYP shows a small red-shift in the OG compared to experiment as the size of the diamondoids is increased. However, TD-PBE0 shows a blue-shift, especially for smaller diamondoids.

It is well-known that approximate DFT functionals suffer from delocalization errors in excited states, which reduce the repulsive Coulomb interaction and hence lead to the underestimation of the band gap.[36] In the opposite fashion, HF orbitals suffer from localization errors, leading to the overestimation of the band gap. The B3LYP hybrid functional, which contains both components, benefits from a cancellation of errors, especially in smaller diamondoids. Since HF localization errors saturate with system size, hybrid functionals show a red shift with increasing size, which is apparent in Table I for B3LYP and TD-B3LYP. TD-B3LYP OGs are in better agreement with experiment than TD-PBE0 for smaller diamondoids. CH$_4$ is one of the species used to parameterize the B3LYP functional, which is therefore expected to perform well for smaller diamondoids. The opposite situation occurs in larger diamondoids, where PBE0-predicted OGs are closer to experiment. This could result from the smaller exact-exchange contribution to the B3LYP functional compared to the PBE0 functional (20% against 25%), giving a better cancellation of errors for smaller diamondoids. With increasing size, the localization error saturates, and the DFT delocalization error is better canceled by the larger exact-exchange term in the PBE0 functional.

Our DMC OGs are greater than the experimental gaps of Landt *et al.* by 0.67 to 0.85 eV. We expect an error of up to about 0.3(1) eV in our DMC-predicted OGs due to the choice of pseudopotentials and the uncertainty in the DFT-optimized geometry, but the overestimation remains significant. The overestimation of the gaps is likely to be a consequence of the fact that the fixed-node approximation retrieves only a finite fraction of the correlation energy. We calculated the HF OG $\Delta E_{HF}$=8.1(2) eV for adamantane using variational quantum Monte Carlo (VMC) without including a Jastrow correlation factor. According to Landt *et al.*, the exact OG of adamantane is $\Delta E = 6.49$ eV, and so the contribution to the gap due to correlation effects is $\Delta E_{corr}$=−1.6 eV. Suppose that our DMC calculations retrieve the same fraction $x$ of the correlation energy in both the ground state and the excited state. Then the DMC gap would be $\Delta E_{DMC}=\Delta E_{HF} + x\Delta E_{corr}$. Since our DMC gap is $\Delta E_{DMC}$=7.35(8) eV, the fraction of correlation energy retrieved would be $x \approx 47\%$, which is implausibly low. The exact fraction of correlation energy retrieved is, of course, unknown, but 90% is a plausible figure based on the performance of DMC with a single-determinant Slater-Jastrow wave function in studies of atoms.[37] Hence it is likely that we retrieve a significantly smaller fraction of the correlation energy in our excited-state calculations than we do in our ground-state calculations.

In another study,[38] DMC overestimated the excitation energies of carbon fullerenes by about 0.8 eV. The authors of Ref. [38] suggested that this demonstrates the need for a more sophisticated wave function. We have found that including backflow correlations does not alter the DMC OG of adamantane significantly. LDA orbitals show degeneracy or near degeneracy in the excited states of all these molecules, especially for adamantane with $T_d$ symmetry. However, using three determinants does not improve the OG of the singlet excitation of adamantane significantly [by only 0.1(1) eV]. This suggests that large numbers of determinants in the trial wave function are required to obtain highly accurate gaps using DMC for symmetrical structures like diamondoids and fullerene derivatives.

The C-H bonds on the surface of diamondoids form surface dipoles. The effects of the interaction of the electrons with these surface dipoles and also with surface phonons on the optical gaps of diamondoids should be investigated in future work.

## B. Optimized geometries, emission gaps and Stokes shifts

The origin of the Stokes shift stems from the rearrangement of the atomic configuration of a molecule in the excited state. Our LDA calculations illustrate that, while the atomic configuration of methane shows a remarkable distortion in the relaxation into point C, the rigid structures of diamondoids resist distortion and rearrangement after excitation in the presence of the electron-hole pair. The structures of methane and adamantane at point A and after global relaxation into point C are shown in Fig. 4. The C-H bond length of methane shows a 7.6% enhancement and its tetrahedral structure in the ground state distorts into a near-flat configuration with broken bonds, resulting in a large reduction in the total energy and therefore EG, leading to a huge Stokes shift of the order of 10 eV. In contrast the change in the tetrahedral angles of adamantane is at most 4º. The presence of degeneracy or near degeneracy in the excited states of diamondoids suggests that these structures should be subject to a Jahn-Teller distortion effect. Indeed the difference of the total energy at point C with and without the relaxation of the symmetry constraint is calculated to be 0.24 eV and 0.4(1) eV using the LDA and DMC methods, respectively. In comparison with the Stokes shift, this amount is nontrivial. An analysis of the DFT orbitals of adamantane in the excited state (point B) shows a three-fold degeneracy for the LUMO and two down-spin occupied orbitals (see Fig. 3). This results in a Jahn-Teller distortion with a splitting of the degeneracy by upwards shifting of the LUMO and downwards shifting of the LUMO-1 and LUMO-2 states.

In Tables II and III we report our EGs and Stokes shifts along with the available experimental data. The Stokes shifts of diamondoids are significantly less than that of methane according to all the theoretical methods. In agreement with experiment, all theoretical methods predict Stokes shifts of less than 1 eV. All theories predict a decrease in the Stokes shift as the size of the diamond NC is increased. Although LDA underestimates the experimental OGs, the LDA Stokes shift of adamantane is close to the experimental value. Unfortunately there are no available experimental data for the Stokes shifts of larger diamondoids, but according to both our DMC and our LDA calculations, the Stokes shifts of diamondoids decrease on average by approximately 0.1 eV per cage, although this rate decreases as the size of the molecule increases.

## C. Self-trapped excitons

Optical absorption in semiconductor NCs leads to the creation of a hole in the bonding state and the excitation of an electron to an antibonding LUMO state, which tends to weaken the bonding. For molecules embedded in an elastic surface medium, the distance between the constituent atoms can increase after excitation. If, following the creation of an exciton in an optical absorption, the length of one particular covalent bond of an NC is stretched beyond a critical value $Q_C$ [see Fig. 1(b)], the exciton can migrate towards this bond and either the electron or the hole can localize over it, forming an STE state.[39]

The signature of an STE that we have looked for in our calculations is the localization of the electron and hole states at point C of Fig. 1 on distorted bonds. In diamondoids, which have rigid internal structures, it may be supposed that an STE is more likely to occur on one of the C-H bonds at the surface, where the elastic response is relatively weak. The lengths of the C-C and C-H bonds in the relaxed excited state relative to the corresponding lengths in the ground state calculated for small diamondoids using the LDA approach are plotted in Fig. 5. For adamantane, the smallest diamondoid, nearly all the bond lengths in the relaxed excited state are distributed around the ground-state values, and the change in length is very small. Most of the C-C bonds are compressed, with only three being stretched. For these three bonds, the bond-length enhancement is only 3.5%, which is small compared with the LDA prediction of a 15% increase in one of the Si-Si bond lengths of $Si_{29}H_{36}$, giving rise to a 2.92 eV Stokes shift.[40] In contrast to the C-C bonds, all the C-H bonds in adamantane are stretched from their ground-state values, but a particular C-H bond is stretched more than the others, by 4.8%. LDA electron and hole orbitals in the excited state are plotted in Fig. 6. The electron orbital of adamantane shows abnormal localization on the most-stretched C-H bond, and is centered over the corresponding H atom. A weaker version of this feature can be observed for the hole state. This suggests an STE is localized upon this particular C-H bond. Photoluminescence emission from an STE in adamantane has been reported in Ref. [33]. However, for diamondoids larger than diamantane this feature begins to weaken. As the size of the particles is increased from adamantane to tetramantane, the localization of the LDA electron orbital over a special bond or its constituent atoms becomes less noticeable, although the tendency of the electron orbital towards the interior of the NC is evident. An analogy between Figs. 5 and 6 may be observed. The electron state localization in Fig. 6(a)–(d) evolves from the outer surface (over the C-H bonds) to the interior (over the C-C bonds). Simultaneously, the region of greatest bond-stretching moves from the C-H bonds on the surface [Fig. 5(a)–(d)] to the C-C bonds in the interior of the NC, suggesting that for larger diamondoids a weak STE occurs over a C-C bond in the interior instead of a C-H bond on the surface. This explains the rapid decrease in the Stokes shift as the diamondoid size increases.

## V. Conclusions

We have calculated and analyzed excitation and emission energies as well as Stokes shifts for small diamondoids using two different theoretical methods: DFT and DMC. In addition, absorption spectra were investigated by TD-DFT using the hybrid B3LYP functional. In general these methods give an accurate description of the optical properties of diamondoids. The irregularities in the trends of optical gaps and absorption spectra with size measured in recent experiments[10] are well reproduced by our TD-B3LYP calculations. Both B3LYP and TD-B3LYP calculations are successful in predicting the OGs of diamondoids, although they (especially TD-B3LYP) show red-shift for larger diamondoids. DMC overestimates the OGs by approximately 0.8(1) eV, which is likely to be due to a difference in the fraction of the correlation energy retrieved in the ground state and excited state in our calculations, resulting in a poor cancellation of errors.

All methods, especially the LDA, can accurately replicate the experimental Stokes shift of adamantane. Our calculations show that the Stokes shift is size-dependent and that it decreases by approximately 0.1 eV per diamond-like cage; however the rate of decrease falls off with increasing size. Although we predict Stokes shifts of up to 1 eV for small diamondoids, this is still small in comparison to Si NCs.[40] Our DFT calculations enable us to deduce that the stiff framework of C-C bonds in the diamond-like structure inhibits relaxation in the excited state, and the largest contribution to dissipation comes from symmetry relaxation. According to our calculations, this symmetry distortion is a Jahn-Teller effect: for example, in adamantane, a three-fold degeneracy in the first excited state is broken, contributing 0.4(1) eV to the Stokes shift. Also, our DFT calculations indicate that an STE is localized upon a single stretched C-H bond in the excited state of adamantane. As the size of the molecule is increased, the largest stretched bond in the excited state moves from the C-H bonds on the surface to the C-C bonds in the interior of the molecule. This is accompanied by the evolution of the electron state towards the C-C bonds in the interior of larger diamondoids. Thus the formation of a weaker STE on C-C bonds in the interior part is more probable than the formation of an STE on the surface.

## Acknowledgements

N.D.D. thanks the Leverhulme Trust for financial support.

TABLE I. Optical gaps calculated using different methods along with available experimental data in eV. The B3LYP and TD-B3LYP gaps were calculated using a 6-31++G(d,p) basis set. For the LDA gaps, a plane-wave basis was used.

[a]Ref. [35]
[b]Ref. [3]
[c]Ref. [34]
[d]Ref. [10]

| Molecule | Formula | Sym. | LDA | B3LYP | TD-B3LYP | TD-PBE0[a] | DMC | DMC[b] | EXP[c] | EXP[d] |
|---|---|---|---|---|---|---|---|---|---|---|
| Methane | $CH_4$ | $T_d$ | 9.46 | 9.87 | | | 10.25(4) | | | |
| Adamantane | $C_{10}H_{16}$ | $T_d$ | 5.76 | 6.48 | 6.47 | 6.66 | 7.35(8) | 7.61(2) | 6.03 | 6.49 |
| Diamantane | $C_{14}H_{20}$ | $D_{3d}$ | 5.43 | 6.30 | 6.57 | 6.75 | 7.07(8) | 7.32(6) | 5.82 | 6.40 |
| Triamantane | $C_{18}H_{24}$ | $C_{2v}$ | 5.21 | 5.97 | 5.93 | 6.12 | 6.80(8) | | 5.68 | 6.06 |
| Tetramantane | $C_{22}H_{28}$ | $C_2$ | 5.05 | 5.88 | 5.83 | 6.01 | 6.76(5) | | 5.60 | 5.95 |
| Pentamantane | $C_{26}H_{32}$ | $T_d$ | | | | 5.86 | | 7.04(6) | 5.51 | 5.81 |

TABLE II. Emission gaps calculated using different methods along with available experimental data in eV. The B3LYP energies were calculated using a 6-31++G(d,p) basis set. For the LDA gaps, a plane-wave basis was used.
[a]Ref. [33]

| Molecule | Formula | LDA | B3LYP | DMC | EXP[a] |
|---|---|---|---|---|---|
| Methane | $CH_4$ | 1.05 | 0.31 | 0.57(6) | |
| Adamantane | $C_{10}H_{16}$ | 4.99 | 5.62 | 6.48(8) | 5.8 |
| Diamantane | $C_{14}H_{20}$ | 4.83 | 5.53 | 6.42(8) | |
| Triamantane | $C_{18}H_{24}$ | 4.72 | 5.44 | 6.40(8) | |
| Tetramantane | $C_{22}H_{28}$ | 4.64 | 5.39 | 6.29(8) | |

TABLE III. Stokes shifts calculated using different methods along with available experimental data in eV. For the B3LYP calculations, a 6-31++G(d,p) basis set was used; for the LDA calculations, a plane-wave basis was used.

[a]Ref. [33]

| Molecule | Formula | LDA | B3LYP | DMC | EXP[a] |
|---|---|---|---|---|---|
| Methane | $CH_4$ | 8.41 | 9.56 | 9.68(7) | |
| Adamantane | $C_{10}H_{16}$ | 0.77 | 0.86 | 0.9(1) | 0.7 |
| Diamantane | $C_{14}H_{20}$ | 0.6 | 0.77 | 0.6(1) | |
| Triamantane | $C_{18}H_{24}$ | 0.49 | 0.53 | 0.4(1) | |
| Tetramantane | $C_{22}H_{28}$ | 0.41 | 0.49 | 0.47(9) | |

FIG. 1. (Color online) (a) Schematic diagram of the ground-state and excited-state singlet and triplet potential energy surfaces of a semiconductor NC in terms of the atomic coordinates ($Q$). Light is absorbed by exciting the NC from the atomic configuration that minimizes the ground-state energy, point A, to the optically active singlet state $B_s$, which has the same geometry as point A. Emission occurs from the relaxed triplet state, point C, leading to a red-shift of the emission line. $E_A$, $E_B$, $E_C$, and $E_D$ are the energies corresponding to points A, B, C, and D, respectively, in the energy surfaces shown in the figure. (b) Schematic diagram in terms of $Q$ for delocalized excitons (DE) and STEs in NCs, showing the corresponding absorption (solid lines) and emission (dashed lines) transitions. If the chemical bond is stretched beyond a critical value $Q_C$, an STE state will be created.

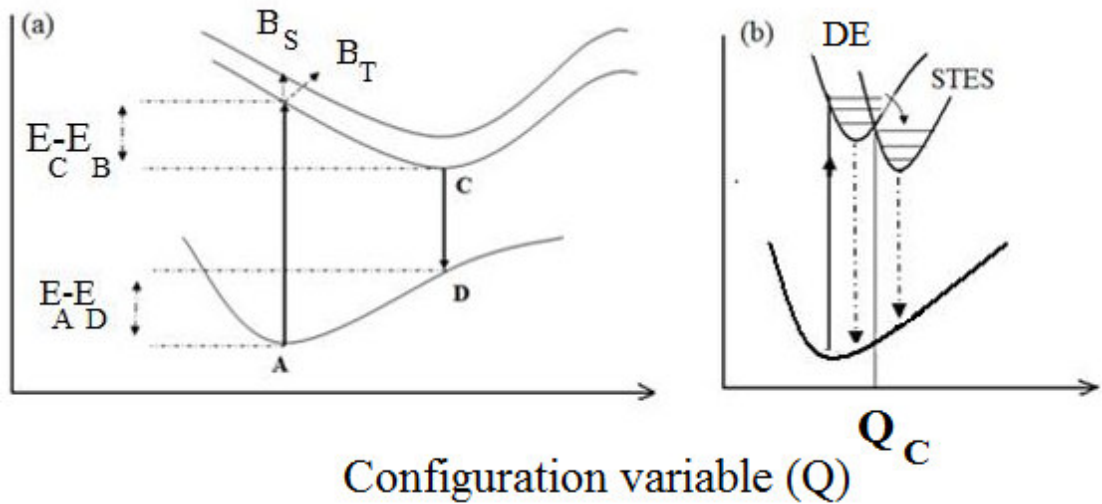

FIG. 2. (Color online) Theoretical excitation spectra from TD-B3LYP calculations for (a) adamantane, (b) diamantane, and (c) triamantane. The calculated lines are Gaussian-broadened by 0.02 eV. Panel (d) shows experimental excitation spectra from Ref. [10]. The structure of each molecule is given above the corresponding spectrum.

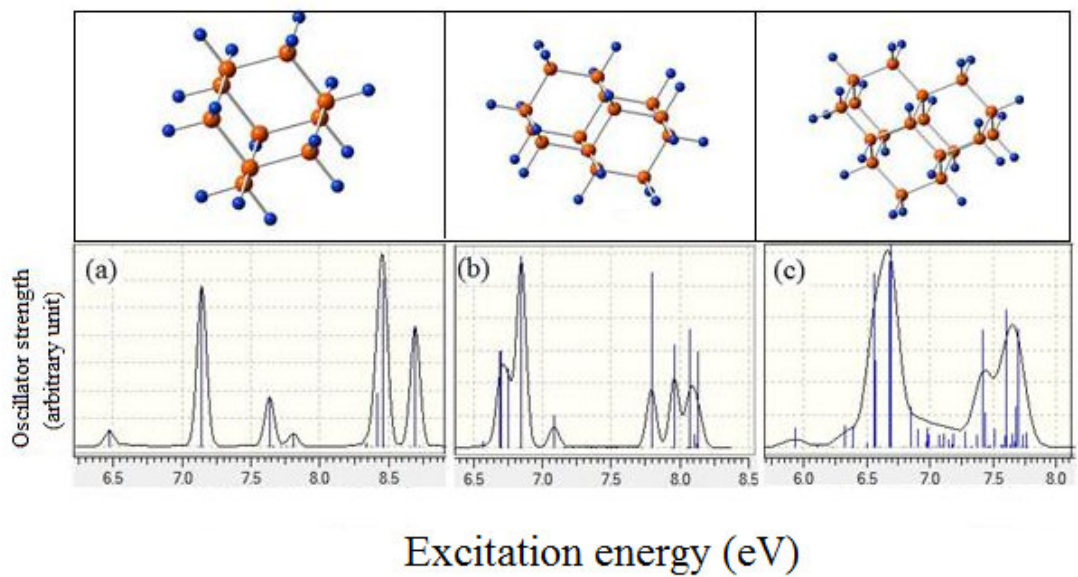

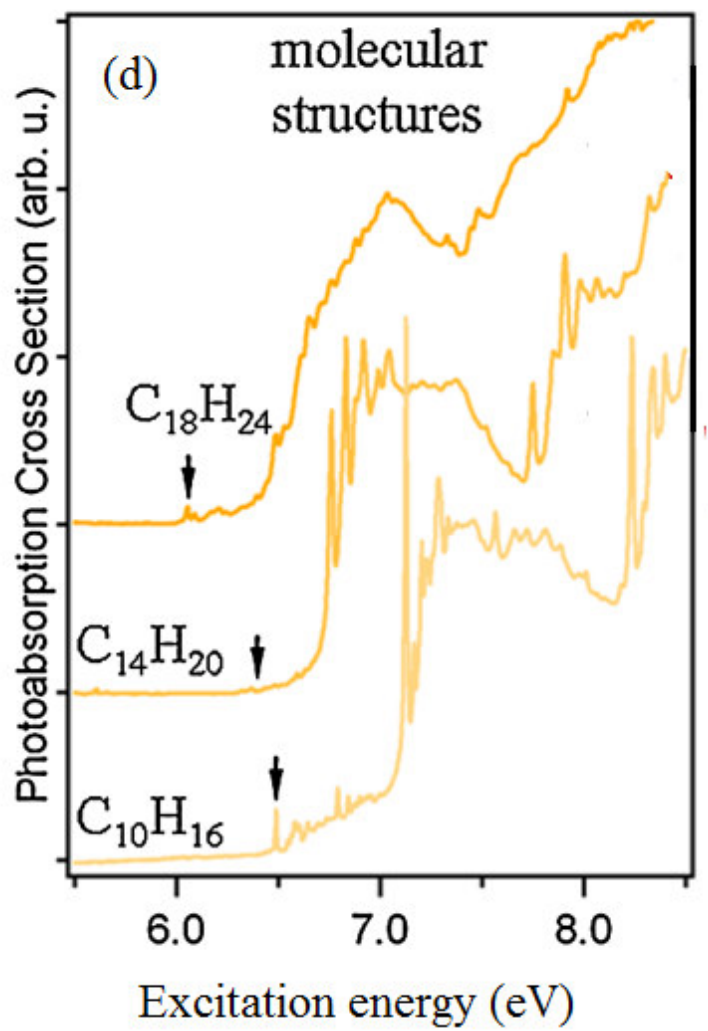

FIG. 3. LDA energy diagram of up- and down-spin electron and hole levels of adamantane in an excited state. The hole is indicated by a small circle. For each level, the energy (in a.u.) and the symmetry of the orbital are shown. Panels (a) and (b) show the energy levels of down- and up-spin electrons at point B. Before the Jahn-Teller relaxation, the hole and two $T_2$-symmetric occupied down-spin levels are degenerate. Panels (c) and (d) show the energy levels of, respectively, up- and down-spin electrons after the symmetry relaxation. The energy levels of the down-spin electrons split, with a reduction in the symmetry of the levels. $\Delta E_{JT}$ is the reduction in the energy of the degenerate levels after the Jahn-Teller relaxation.

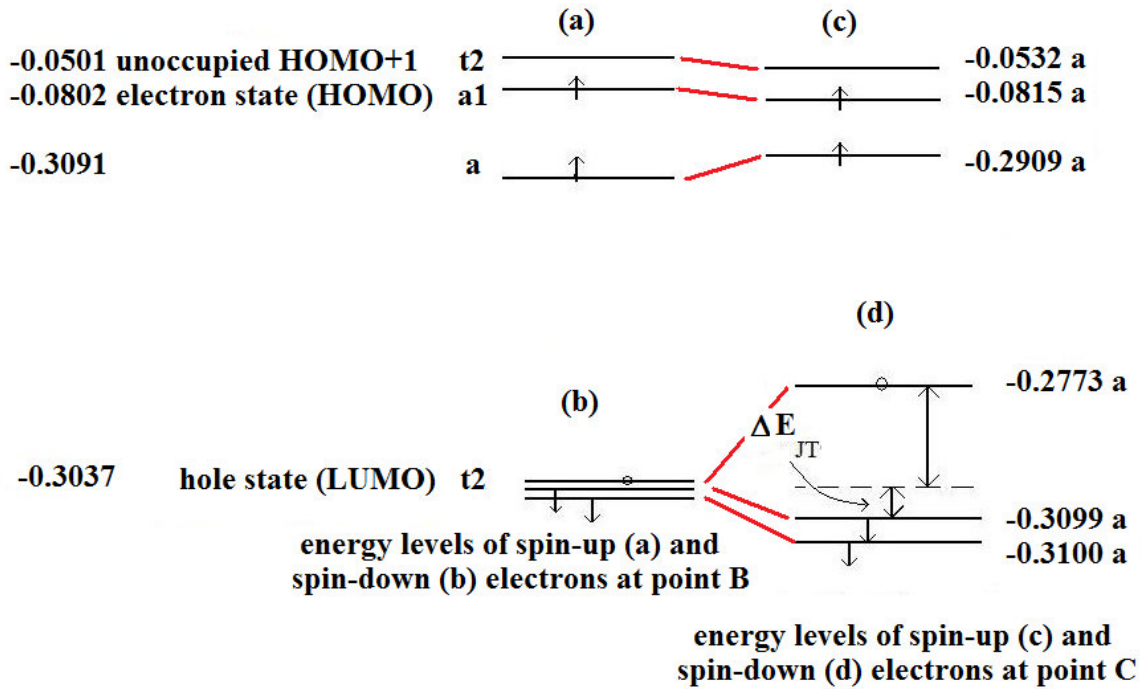

FIG. 4. (Color online) The optimized geometries of methane and adamantane in (a) the ground state and (b) the geometry- and symmetry-relaxed excited state, from LDA calculations using plane-wave basis sets. All lengths are in Å and angles are in degrees.

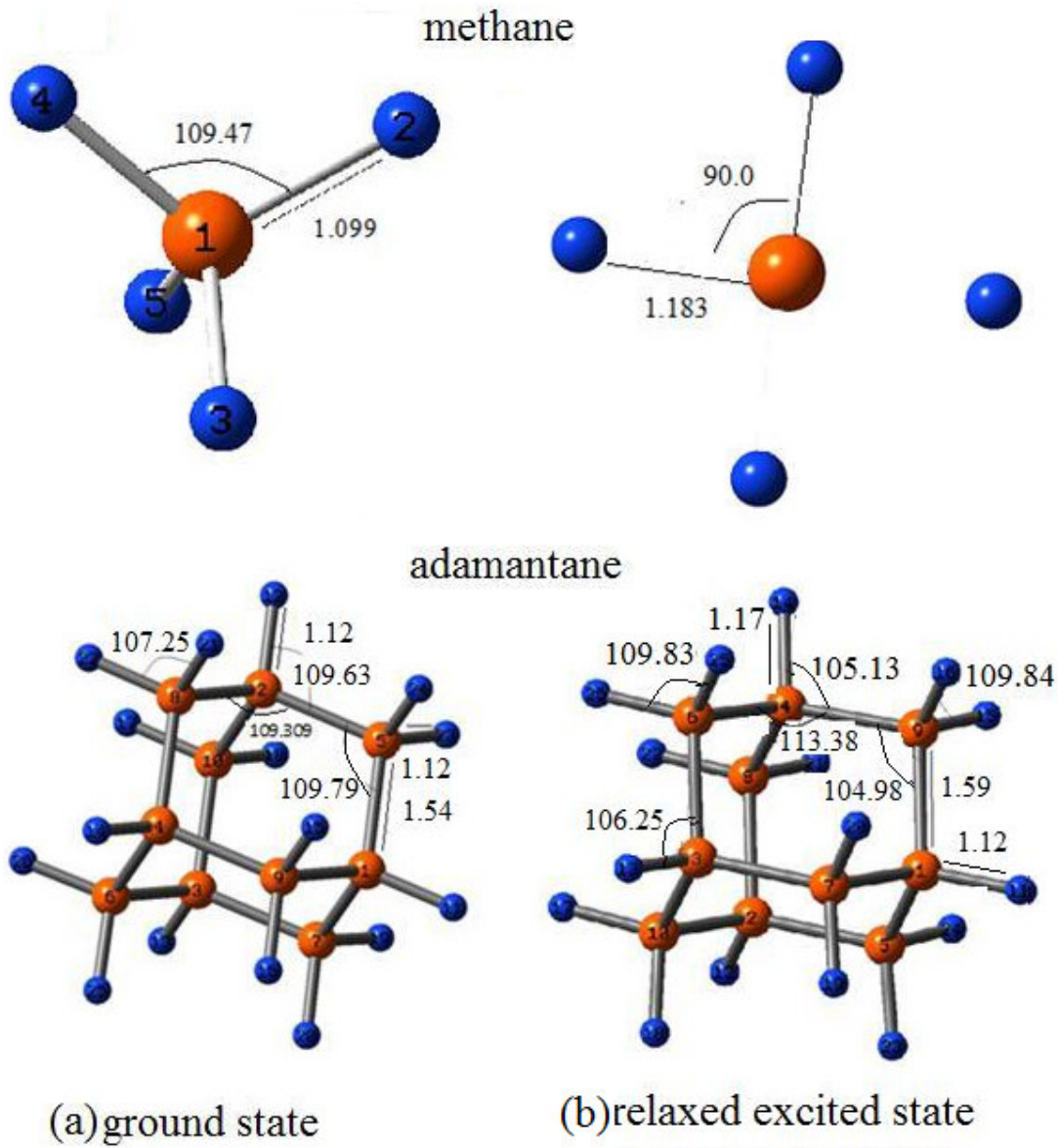

(a) ground state

(b) relaxed excited state

FIG. 5. (Color online) Distribution of the C-C (left panel) and C-H (right panel) bond lengths in the relaxed triplet excited states relative to their ground-state values: $[L(\text{C-C})_{ES}-L(\text{C-C})_{GS}]/L(\text{C-C})_{GS}$ and $[L(\text{C-H})_{ES}-L(\text{C-H})_{GS}]/L(\text{C-H})_{GS}$. $L(\text{C-C})_{ES}$ and $L(\text{C-C})_{GS}$ denote the C-C bond lengths in the excited and ground states, respectively. $L(\text{C-H})_{ES}$ and $L(\text{C-H})_{GS}$ denote the C-H bond lengths in the excited and ground states, respectively. From top to bottom the panels correspond to (a) adamantane, (b) diamantane, (c) triamantane, and (d) tetramantane. The lines are artificially Gaussian broadened by between 0.5 and 0.05. According to (d), nearly no change in the C-H bond length is observed in the excited state in comparision with the ground state.

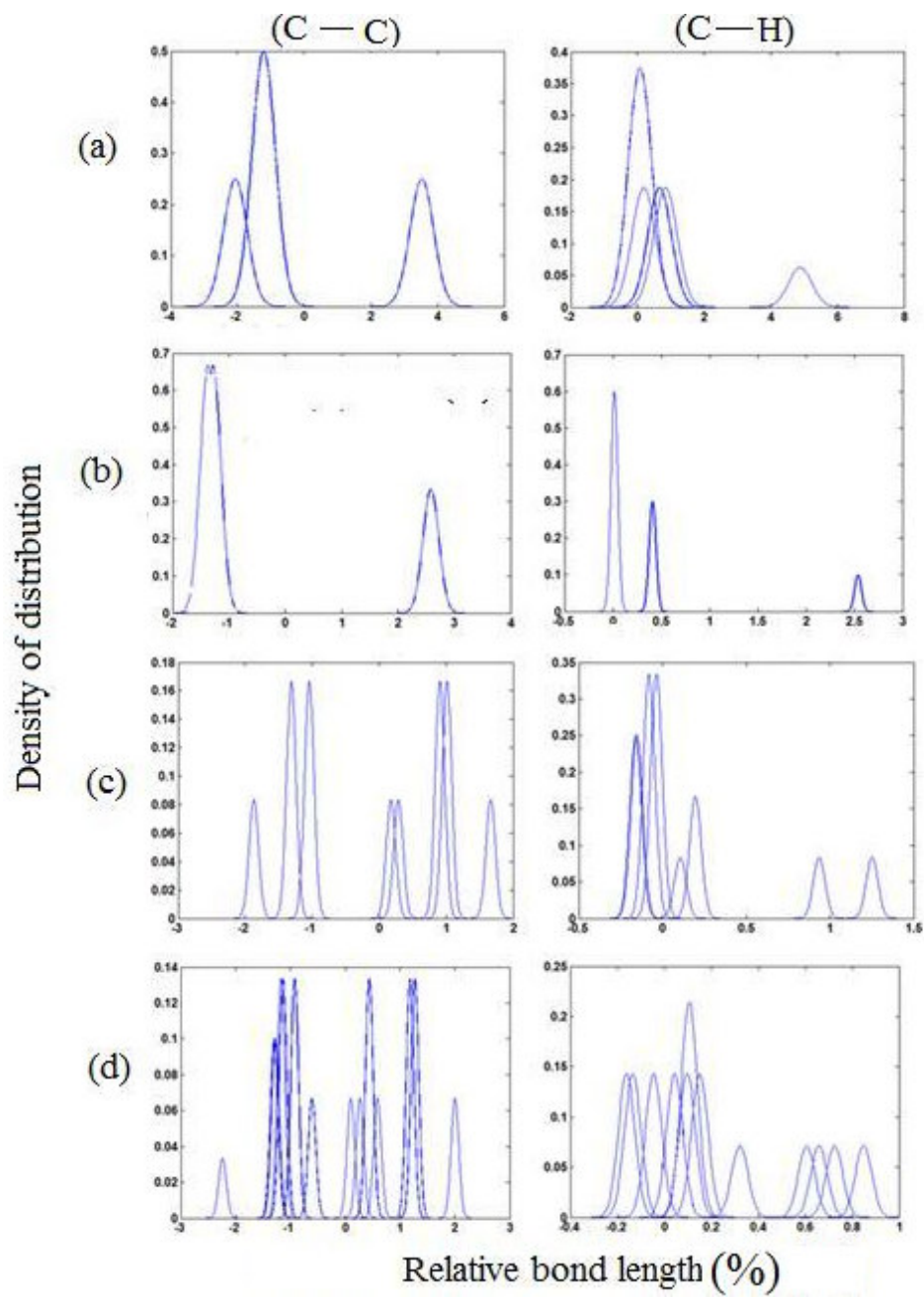

FIG. 6. (Color online) Isosurfaces of charge density of the electron and hole orbitals with isodensity value 0.02 a.u. in the triplet excited-state configuration. The arrow indicates the most-stretched C-H bond of adamantane, on which the electron state localizes. Panels (a), (b), (c), and (d) show adamantane, diamantane, triamantane, and tetramantane, respectively.

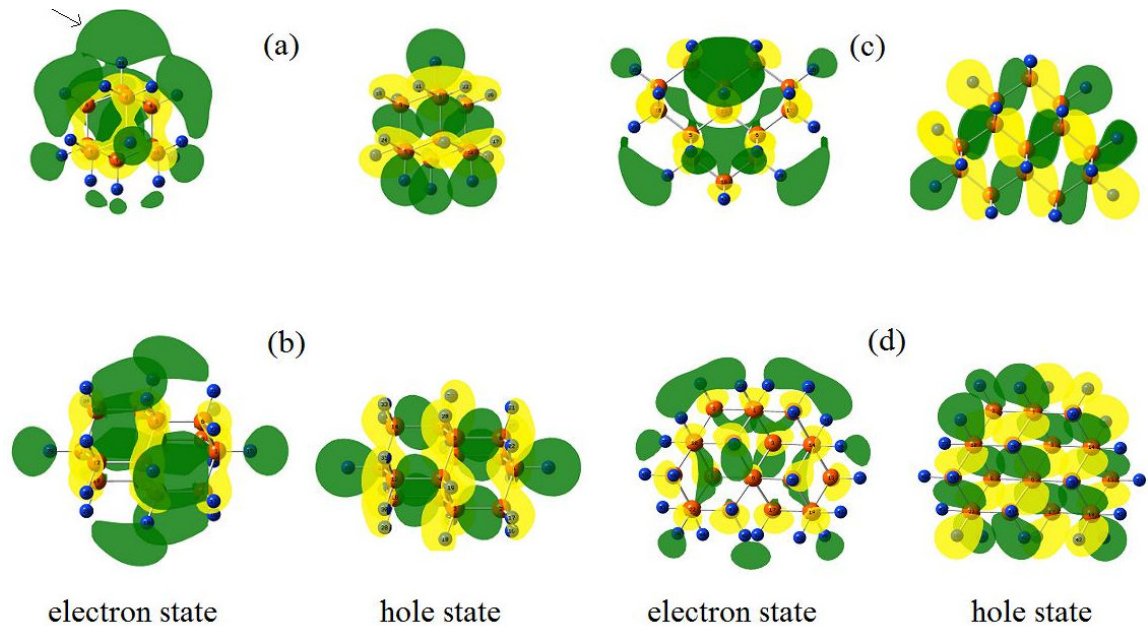